# Sub-10-micron thick Ge thin films from bulk-Ge substrates via a wet etching method

Liming Wang and Guangrui (Maggie) Xia[1,*]

[1] Department of Materials Engineering, the University of British Columbia, Vancouver BC, V6T 1Z4 Canada

**Abstract**

Low-defect-density Ge thin films are critical in Ge based optical devices (optical detectors, LEDs and Lasers) integrated with Si electronic devices for low-cost, highly integrated photonic circuits. In this work, Ge thin films prepared by wet etching with four different solutions were studied in terms of the surface morphology, defect density and achievable thickness. Both nanostrip-based solution (1:1:10) and HCl-based solution (1:1:5) were able to wet-etch 535 µm thick bulk-Ge substrates to Ge films thinner than 10 µm within 53 hours. The corresponding RMS surface roughness was 32 nm for the nanostrip-based solution and 10 nm for the HCl-based solution. The good quality of bulk-Ge was preserved before and after the etching process according to the HRXRD results. The low threading dislocation density of 6000-7000 $cm^{-2}$ was maintained in the process of wet etching without introducing extra defects. This approach provides an inexpensive and convenient way to prepare sub-10-micron thick Ge thin films, enabling future studies of low-defect-density Ge-based devices such as photodetectors, LEDs, and lasers.

Highlights:

(1) Two recipes with reasonable post-etching morphology for Ge were developed

(2) A 9.2 µm thick Ge thin film and a 4.1 µm thick Ge thin film were successfully thinned down from a 535 µm thick bulk-Ge substrate

(3) Low threading dislocation density of 6000-7000 $cm^{-2}$ maintained before and after etching process

(4) Good crystal quality was reserved during the etching

**Introduction**

Practical Si-compatible light sources especially Si-compatible lasers have been sought for decades for applications in silicon photonics.[1-3] While early attempts to integrate III-V semiconductor lasers on a Si platform has made significant progress with quantum dot (QD) lasers on Si or Ge[4-6], the co-processing of III-V semiconductors in Si-based fabrication facilities has a prohibiting auto-doping problem, as Si and III-V semiconductor elements are dopants to each other. This has significantly limited the manufacturing and adoption of Si-compatible QD lasers.

Ge is drawing growing interest due to its CMOS compatibility and potential of light emission.[7-9] Even though Ge has an indirect bandgap, the difference between the direct band gap at the Γ valley and the indirect band gap at the L valley is only 136 meV at room temperature.[10] Methods to turn Ge into a pseudo direct bandgap material include tensile strain and heavy n-type doping, where tensile strain decreases the difference between the L valleys and the Γ valley, and the electrons from n-doping fill the L valleys to the level of the Γ valley bottom to compensate for the remaining energy difference.[11] The first optically pumped Ge laser and electrically pumped Ge laser were demonstrated using n-doping and thermal tensile strains of about 0.2% from thermal expansion mismatch between Ge and Si.[12-13] However, those epitaxial Ge-on-Si lasers had very low efficiencies and high threshold currents due to the high threading dislocation density (TDD) in epitaxial Ge on Si, which was in the range of $10^8$-$10^{10}$ cm$^{-2}$, a few orders of magnitude larger than the TDD of bulk-Ge wafers, commonly ⩽ $10^4$ cm$^{-2}$ and even dislocation free.[14-16] Such high threading dislocation density of Ge increases the Shockley-Read-Hall (SRH) recombination rate which lowers the carrier lifetime leading to the poor performance of Ge-based optoelectronic device. What is the ultimate performance potential that Ge lasers can reach? Although several theoretical papers have predicted that the performance will improve dramatically with high quality Ge, experimental studies are still lacking.[15, 17] Our aim is to answer the question of Ge laser potential experimentally using the highest quality Ge, bulk-Ge, which has not been studied for Ge laser related applications before. To make Ge lasers from bulk-Ge, the first step is to obtain Ge thin films of micron scale from bulk-Ge wafers of a few hundred micron thick, which is the goal of this work. Besides Ge lasers, Ge thin films were also used in photodetectors and LEDs.[18-20] Ge thin films with thicknesses of several micrometers are of great interest for solar cell application but have not yet been accomplished to the best of our knowledge.[21-23] Meanwhile, Ge thin films have been demonstrated to be a good model for introducing large mechanical tensile strain to achieve a direct band to band emission.[24-26]

There are two strategies to prepare Ge thin films: a bottom-up approach and a top-down approach. The bottom-up approach is to use tools like chemical vapor deposition (CVD) to deposit a thin layer of Ge on a substrate like Si, then the substrate is removed leaving the Ge thin film.[24] Due to the fact that the Ge thickness can be well controlled with the deposition rate and time, this method can prepare Ge thin film with desired thickness.[27-28] One problem for this method is that the thickness is limited ($\leqslant 2000$ nm[29]) as the isolated crack will show up with increasing thickness due to the thermal strain.[30] And owing to the lattice mismatch between the Si substrate and epitaxial Ge, the epitaxial Ge has a threading dislocation density as high as $10^8$ cm$^{-2}$ which impede the performance.[29] Preparing a Ge thin film from a low defective bulk-Ge is a top-down approach. While method like smart cut was propose to obtain a Ge thin film on Si substrate[16, 31-32], solution-based methods like wet etching to thin Ge are much cheaper and more accessible to get Ge thin film especially in the early R&D stage. With a low defective bulk-Ge, it is possible to get a low defective Ge thin film for potential optoelectronic applications.

As the pioneering transistor material, Ge's first wet etching study dates back to the year of 1955, when Paul. R. Camp studied the etching rates of Ge with solutions composed of $H_2O_2$, HF and water as a function of etchant composition and crystal orientation and impurity.[33] More etchants for Ge wet etching were studied in the following years, and the related etching rates of Ge for different solutions were well summarized in literatures.[34-35] However, there are only limited reports on the preparation of Ge thin film from bulk-Ge via wet etching. Two literatures in the hope to get a thin film with a thickness of 5 - 10 µm ending up with thicknesses of $\geqslant 80$ µm.[21, 23] Another literature managed to get a thin film with a thickness of 28 µm but the overall morphology after the etching process was missing.[26]

Considering the fact that the surface morphology plays an important role in optoelectronic devices, in this work, we investigated the post-etching Ge morphologies resulted from different etching recipes with an optical microscope and a 3D optical profilometer. The optical microscope instead of SEM was used here because a larger area morphology was preferred to represent the overall flatness. And the 3D optical profilometer is a good method for quantitatively measuring roughness for a larger area ($\geqslant$ 300 µm × 300 µm) comparing to AFM (mostly $\leqslant$ 30 µm × 30 µm). Two recipes with reasonable post-etching morphology were developed to get sub-10 µm thick free-standing Ge thin films from a 535 µm-thick Ge wafer.

**Experiment method and design**

The beginning substrate is a 4-inch n-type (0.173-0.25 Ohms*cm at 295 K) double sides polished (100) Ge Czochralski wafer, that was obtained commercially. The wafer was diced into 1 cm × 1 cm size wafer pieces before the etching process. The initial thickness was 535 µm, which was measured by a micrometer gauge and an optical microscope, as shown in Figure S1. All the Ge pieces were cleaned sequentially with acetone, isopropyl, and DI water, and dried with $N_2$ gas. All the wet etching was conducted in a wet bench with good ventilation at an ISO-7 class cleanroom with the temperature well controlled at 22 °C.

Based on literature studies, our work started with four different solutions as discussed below. Acid + $H_2O_2$ solutions are widely used in Ge etching and fab compatible as the etchants, and were chosen to oxidize Ge step by step into germanic acid, $Ge(OH)_4$, and dissolve the products.[36] In this work, $H_2SO_4$, HCl and HF-based acids were chosen as the three acid options. In cleanrooms, nanostrip solution (Nano-strip 2X: 85% $H_2SO_4$, ≤ 1% $H_2O_2$) is more frequently used than concentrated $H_2SO_4$ solution (96%), we took nanostrip solution instead of concentrated $H_2SO_4$. To simplify the description, the solution with HCl solution (37 %) as the acid named the HCl-based solution (X: Y: Z) where X: Y: Z was used to indicate the volume ratio of HCl solution (37 %), $H_2O_2$ solution (30%) and DI water, so was for nanostrip-based solution (X:Y:Z)(Nanostrip solution 2X which contains 85% $H_2SO_4$ and ≤ 1% $H_2O_2$ as acid), HF-based solution (X:Y:Z) (49 % HF solution as acid) and $HNO_3$-based solution (X:Y:Z)(70 % $HNO_3$ solution as acid). One more solution selected as the control solution was the $HNO_3$-based solution (1:0:1) which was reported by the literature for Ge etching.[37]

Initially, each Ge piece (1cm × 1cm × 535 μm) was placed on the bottom of a beaker, etchant with a volume of 35 mL was added into the beaker for the etching processes (24 h etching). After etching, all samples were cleaned with DI water and dried with $N_2$ gas before further characterizations. Optical images of the surface morphologies were taken with a Nikon ECLIPSE LV150 optical microscope. To get a thin film, more etching time was required. It was noticed that some sediment could be seen on the top of the etched Ge after long etching process (before cleaning). To exclude the influence of the potential sediment during long time etching and speed up the etching processes, Ge was placed vertically in the beaker on a small Teflon stand with double sides being etching at the same time. The post-etching thicknesses for the thin films were measured under the optical microscope. The surface morphology was further evaluated with a 3D optical profilometer (phase shift interferometry). Etching pit density (EPD) measurements were performed to obtain the TDDs before and after etching using an etching solution from literature.[38] Optical microscope was used to observe, count the etch pits and calculate the etch pit density (EPD) with more than three positions being checked. The EPD etchant which was a mixture of 100 mL $CH_3COOH$ (⩾99%), 40 mL 70% $HNO_3$, 10 mL 49% HF, and 30 mg $I_2$ (⩾99.99%) was selected according to the literature [39].

**Results and discussion**

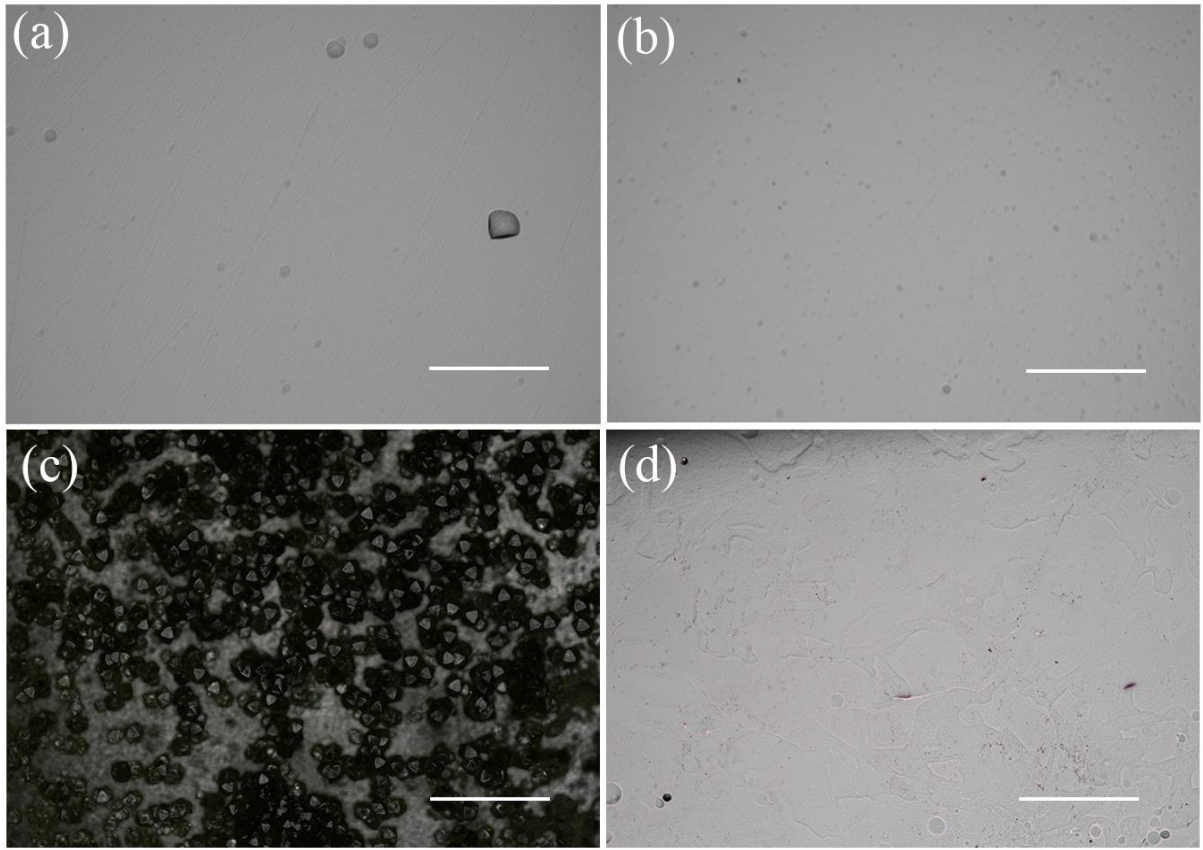

**Figure 1.** 24 h etching in (a) HCl-based solution (1:1:10) (b) Nanostrip-based solution (1:1:10) (c) HNO$_3$-based solution (1:0:1), 25 min etching in (d) HF-based solution (1:1:1), scale bar = 500 µm.

After 24 h etching, the surface morphologies were checked with an optical microscope. As shown in figure 1a and 1b, HCl-based solution (1:1:10) and nanostrip-based solution (1:1:10) etched surfaces were flat with some etching pits. However, black particles/bumps were shown on the surface etched with HNO$_3$-based solution (1:0:1) solution, which could be oxides due to strong oxidizing leaving a rough morphology (figure 2c). The surface height variation of HNO$_3$-based solution (1:0:1) etched surface was further characterized by a stylus profilometer (figure S2) which confirmed large particles with a height of approximately 80 µm were shown on the surface. HF-based solution was firstly tried with ratio of 1:1:1 because it was reported to obtain a 28 µm thick Ge thin film from bulk-Ge.[26] Owing to the fact that HF-based solution had a very high etching rate, Ge sample was totally etched off after 24 h etching. The result for 25 min HF-based solution (1:1:1) etching was shown in figure 2d that the surface was totally roughened by HF-based solution etching. The remaining thickness after 25 min etching was 330 µm which represented that approximately 200 µm was etched off in 25 min (figure

S3d). More experiments were also performed for HF-based solutions with 1:1:1 to 1:1:5 to 1:1:10 volume ratios. As shown in figure S3, with the ratio changing from 1:1:1 to 1:1:5 to 1:1:10, the surface roughness could be improved. However, when the etching time was extended from 25 min to 4h, the surface which could be seen from figure S4 was ruined with obvious cracks on the surface. Therefore, $HNO_3$-based solution and HF-based solutions were eliminated in the further studies, and we focused on HCl and nanostrip-based solutions.

To further study the details of nanostrip-based solutions, nanostrip-based solutions with different volume ratios were prepared. On the high $H_2SO_4$ limit, nanostrip-based solution (1:0:0), no extra $H_2O_2$ solution and DI water, etched Ge sample (figure 2b) showed a flat surface with only a few straight lines on the surface. But it had a low etching rate without obvious change of the thickness after 24 h etching, indicating Ge was rarely etched. This was also observed by other researchers and could be attributed to the strong oxidizing from the high concentration $H_2SO_4$ which created enough Ge oxides (GeO and $GeO_2$) that covered the entire germanium surface to produce passivation and prevent further etching of germanium.[37, 40] On the zero $H_2SO_4$ limit, nanostrip-based solution (0:1:0) etching resulted in a rough Ge surface (figure 3c) with a lot of square voids (figure S5). With a nanostrip-based solution (1:1:0), the etched surface (figure 3d) was getting dark and rough in the optical microscopy, and some black particles appeared on the etched surface. With the ratio change into 1:1:1, the surface showed the similar behavior of $HNO_3$-based solution (1:0:1) indicating a strong oxidative effect on the surface with potential oxidized particles on the surface. The etched Ge samples with the best surface qualities judging from the optical microscopy were the ones etched with a nanostrip-based solution (1:1:5) or a nanostrip-based (1:1:10) solution. Surface etched with mixed with nanostrip-based solution (1:1:5) showed a flat surface with minor holes on the surface. One problem for nanostrip-based solution (1:1:5) solution was that the holes after etching would grow making the surface much rougher (figure S6). And the hole seemed to be located on the location showing the preferred etching for the dislocation positions. For nanostrip-based solution (1:1:20), the surface became rougher with obvious bumps (figure 2h). According to these facts, nanostrip-based solution with a volume ratio of 1:1:10 was used for further etching experiment.

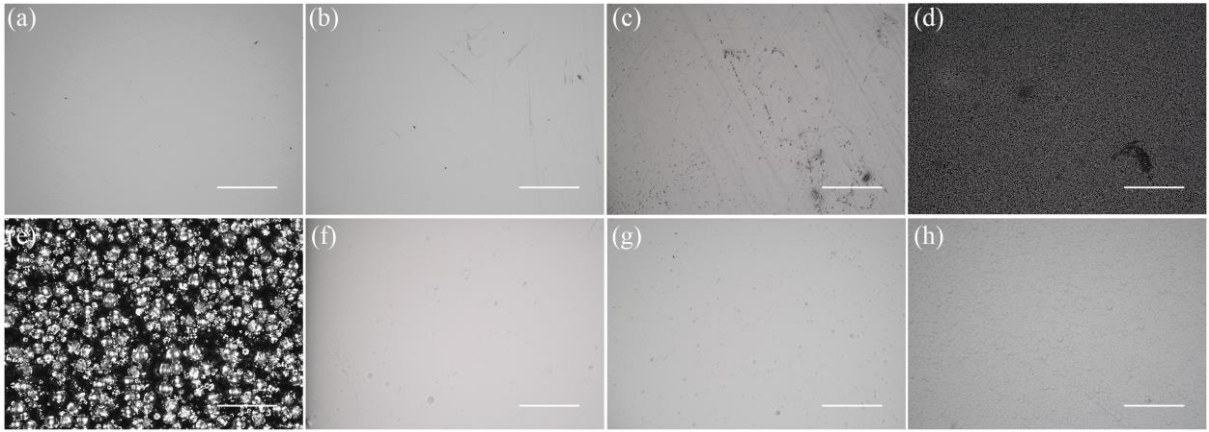

**Figure 2.** Optical images of Ge surfaces: (a) before any etching, after 24 h etching in (b) nanostrip, (c) $H_2O_2$ solution, (d) nanostrip-based solution (1:1:0), (e) nanostrip-based solution (1:1:1), (f) nanostrip-based solution (1:1:5), (g) nanostrip-based solution (1:1:10) and (h) nanostrip-based solution (1:1:20). The scale bar is 500 μm.

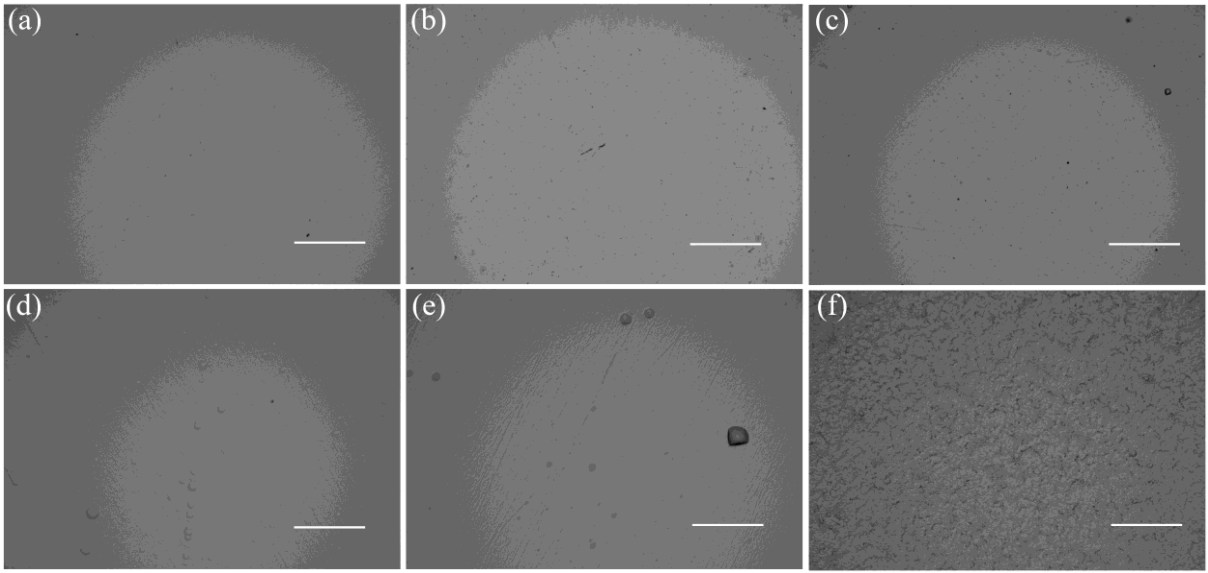

**Figure 3.** Optical images of Ge surfaces: (a) Unetched Ge, 24 h etching in (b) HCl-based solution (1:0:0) (c) HCl-based solution (1:1:1) (d) HCl-based solution (1:1:5) (e) HCl-based solution (1:1:10) (f) HCl-based solution (1:1:20), the scale bar is 500 μm.

HCl-based solutions with different ratios were also explored to optimize the solution ratio. HCl-based solution (1:0:0) etched Ge sample (figure 3b) showed numerous etching pit on the surface. With the ratios changing from the 1:1:1 to 1:1:5, the surface improved. However, when the ratios increased to 1:1:10, the etching pit size increased. The surface etched by HCl-based solution (1:1:20) became quite rough with a mat like surface from the optical microscope.-based on these observations, nanostrip-based solution (1:1:10) and HCl-based solution (1:1:5) were applied here for thin films preparation.

The final results of the thin films were shown in figure 4, both nanostrip-based solution (1:1:10) and HCl-based solution (1:1:5) obtained thin films with thickness of < 10 µm. As shown in figure 4a, Ge etched by nanostrip-based solution (1:1:10) for 57 h demonstrated a thickness of 9.2 µm. The picture of the samples was shown on the top right. Only half of the Ge piece was etched with another half being protected by photoresist during the etching process. The thickness of Ge after HCl-based solution (1:1:5) was 4.1 micrometers (figure 4b). A mirror like surface was still kept for HCl-based solution etched sample with the reflection of a tweezer could be clearly seen (figure 4c). The reflectance before and after the etching were shown in figure 4d, where the unetched Ge had a reflectance of 50 % and the etched Ge had a reflectance of 35% (comparing to standard Si). Overall, approximately 70 % of reflectance was preserved comparing to unetched Ge.

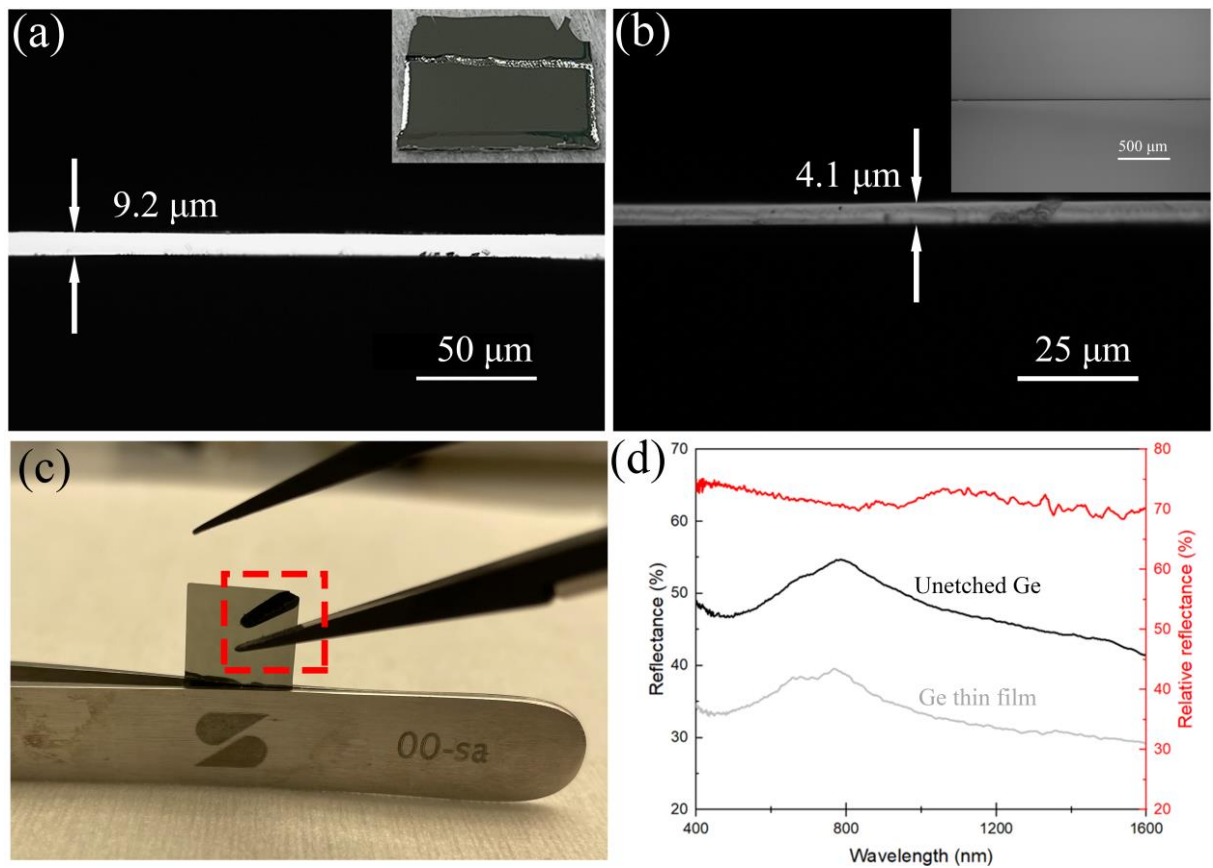

**Figure 4.** (a) Thickness measurement from the top of the Nanostrip-based solution (1:1:10) 57 h etched Ge (b) Thickness measurement from the top of the HCl-based solution (1:1:5) 53h etched Ge (c) Front view of the HCl-based solution etched Ge, and the reflection of a tweezer (red square circled region) (d) Reflectance and relative reflectance of unetched Ge and HCl-based solution (1:1:5) etched 53h Ge thin film

To further quantitatively describe the surface roughness after the etching processes, the surface roughness of unetched Ge and Ge thin films were measured with a 3D optical profilometer. The optical image before the wet etching process was shown in figure 5a, the related surface roughness measured (figure 5e) indicated that the unetched Ge had a roughness of approximately 1.6 nm with some minor polishing traces on the top. After nanostrip-based solution etching for 57 h, the optical images showed a lot of hemi-spherical holes on the top (figure 5b), the surface roughness (figure 5f) increased to 60 nm with holes of different sizes on the surface. This could be improved by an agitation (300 rpm) during the etching process where the surface etching hole sizes decreased (figure 5c) and the surface roughness (figure 5g) dropped to 32 nm. HCl-based solution etched thin film had fewer etching holes and a flatter surface judging from optical microscope (figure 5d), and the surface roughness (figure 5h) was approximately 10 nm which was much better than the nanostrip-based solution etched samples. This also explained why the HCl etched surface had a good reflectance. It could also be expected that the surface roughness of the thin film could be lower starting with a Ge wafer with lower roughness (well polished) and less thickness.

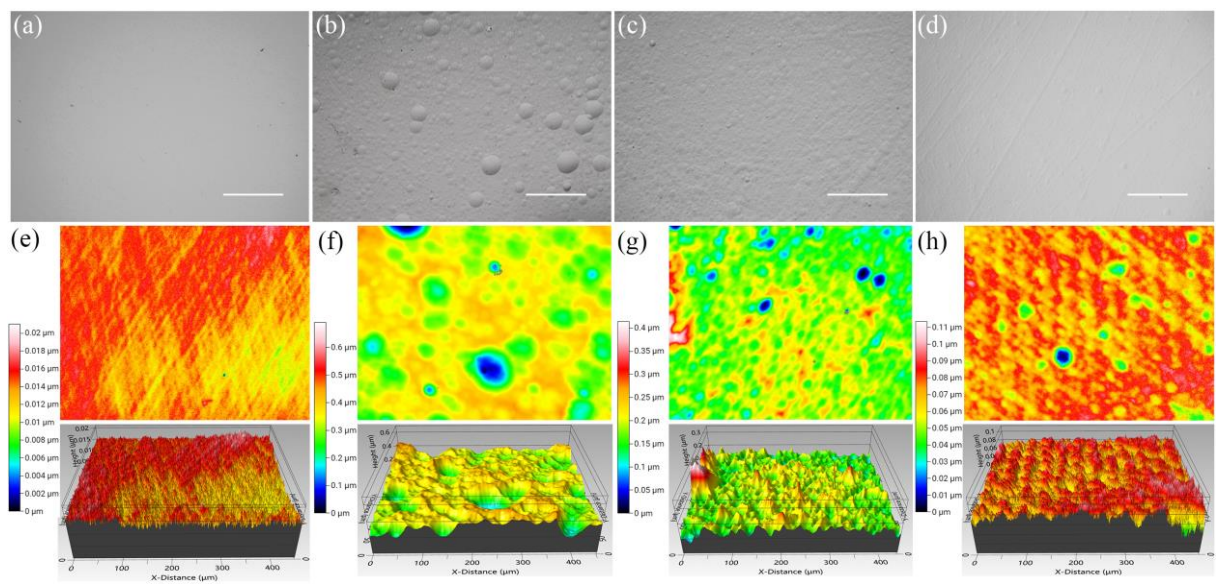

**Figure 5.** Optical images for (a) unetched sample (b) nanostrip-based solution (1:1:10) etched 57 h sample (c) nanostrip-based solution (1:1:10) etched 51 h sample with agitation (d) HCl-based solution (1:1:5) etched 53 h sample without agitation. White light interferometer (Phase shift interferometry) results of for (e) unetched sample, Rsq =1.6 nm (f) nanostrip-based solution (1:1:10) etched sample, Rsq = 60 nm(g) nanostrip-based solution (1:1:10) etched sample with agitation, Rsq = 32 nm (h) HCl-based solution (1:1:5) etched sample without agitation, Rsq = 10 nm

Comparing with the bottom-up approach, one advantage for the top-down approach is the preservation of the good crystal quality due to the fact that the etching process only occur on the surface without damaging the crystalline quality. The crystal quality was demonstrated by HRXRD as shown in figure 6a, both the unetched and the HCl-based solution (1:1:5) 53 h etched thin film had sharp peak for the XRD which indicated a good crystalline quality. And the peak position was exactly the same before and after the etching process which also demonstrated no strain or obvious lattice damage introduced for the Ge thin film. And the threading dislocation density before and after the etching processes were also checked with an etching pit density experiment, as shown in figure 6b and 6c. Owing to the fact that bulk-Ge has a much lower threading dislocation density, it could be difficult to find enough dislocation sites under high magnification. Assuming the bulk-Ge had a dislocation density of $10^4/cm^2$, under a high magnification of 500X, the region under optical microscope is around $4.44 \times 10^{-4}$ $cm^{-2}$ (250 μm × 177 μm), that means averagely only 4 etching pit can be seen in the view which increases the error. Thus, the etching time externed to 90s instead 15s to get obvious etching pit for a large area. The etching pit densities before and after the etching processes remained almost the same level of 6000-7000 $cm^{-2}$ which confirmed the low dislocation density for the wet etched Ge thin film.

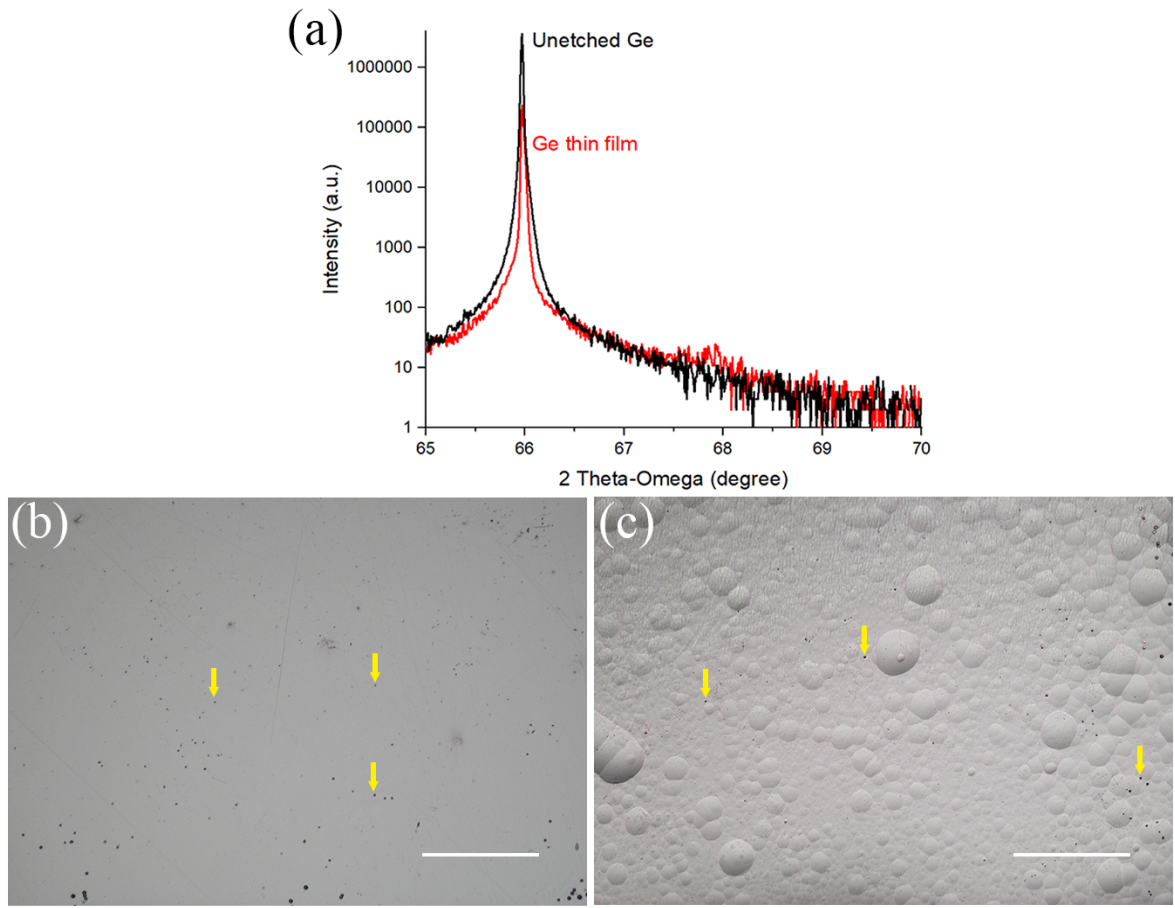

**Figure 6.** (a) HRXRD rocking curves of unetched bulk-Ge and HCl-based solution (1:1:5) 53 h etched thin film, EPD results for (b) unetched sample (c) nanostrip-based solution (1:1:10) 57 h etched sample, the yellow arrow points out the represent of the etching pit, the scale bar is 500 µm.

Even though we have obtained Ge films thinner than 100 µm from wet etching (figure 7a), one problem for wet etching was that the etchant would attack the sample from all directions due to the isotropic etching behavior. As shown in figure 7b, the Ge thin film has a wedge shape. On the edge, the thickness is < 10 µm, but for the middle, the thickness is around 70 µm. The thickness increases by 14.8 µm for every 500 µm length, which translates to an angle of 1.65 degree between the Ge top and bottom surfaces, which is insignificant for device fabrication. This shape was observed in Ge films etched by nanostrip and HCl-based solutions. This shape is a result of three-sided etching at the edge and the isotropic etching (figure 7c).

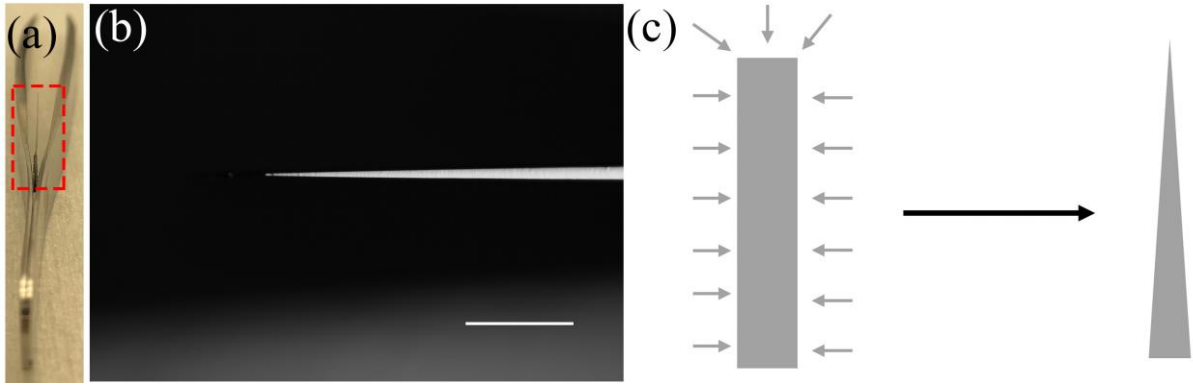

**Figure 7.** (a) Side view of the HCl-based solution (1:1:5) 53 h etched thin film (b) Optical image of the side view of the thin film, scale bar = 500 µm (c) the possible mechanism

**Future works**

Ge will be bonded on a substrate and undergo the wet etching thinning process. A polishing process may also be applied for a bonded Ge thin film on a handle substrate to obtain a lower surface roughness for the future device (LEDs and Lasers) fabrication.

**Conclusion**

In this work, Ge thin films prepared by wet etching with four different solutions were investigated in terms of the surface morphology, defect density, etch rate and achievable thickness. Both nanostrip-based solution (1:1:10) and HCl-based solution (1:1:5) were able to wet-etch 535 µm thick bulk-Ge substrates to Ge films thinner than 10 µm within 53 hours. The corresponding RMS surface roughness was 32 nm for the nanostrip-based solution and 10 nm for the HCl-based solution. The good quality of bulk-Ge was preserved before and after the etching process according to the HRXRD results. The low threading dislocation density of 6000-7000 cm$^{-2}$ was maintained in the process of wet etching without introducing extra defects. This approach provides an inexpensive and convenient way to prepare sub-10-micron thick Ge thin films, enabling future studies of low-defect-density Ge-based devices such as photodetectors, LEDs, and lasers.

**Acknowledgement**

CMC Microsystems, Canada's National Design Network (CNDN), Natural Sciences and Engineering Research Council (NSREC) are acknowledged for providing financial support of the lab work. We would also like to thank Dr. Andrey Blednov and Dr. Mario Beaudoin from UBC nanofab for cleanroom equipment training and assistance, Dr. Qiong Wang from UBC Materials Engineering Department and Dr. Saeid Soltanian from UBC CFET for the help of 3D optical profilometer. Liming Wang acknowledges UBC Four Year Doctoral Fellowship for the financial support.